\documentclass{article}
\usepackage{spconf,amsmath,graphicx,multirow,booktabs,bm,amssymb}
\usepackage{url}

\title{ADD 2022: the First Audio Deep Synthesis Detection Challenge}
\name{{Jiangyan Yi$^{1}$, Ruibo Fu$^{1}$, Jianhua Tao$^{1,2,*}$$\thanks{*Corresponding author.}$, Shuai Nie$^{1}$, Haoxin Ma$^{1,2}$, Chenglong Wang$^{1}$,} \\ \textit{Tao Wang$^{1,2}$, Zhengkun Tian$^{1,2}$, Xiaohui Zhang$^{1}$, Ye Bai$^{1,2}$, Cunhang Fan$^{1,2}$, Shan Liang$^{1}$, Shiming Wang$^{1}$,} \\ \textit{Shuai Zhang$^{1,2}$,Xinrui Yan$^{2}$, Le Xu$^{2}$, Zhengqi Wen$^{1}$, Haizhou Li$^{3}$, Zheng Lian$^{1}$, Bin Liu$^{1}$}}
\address{$^1$National Laboratory of Pattern Recognition, Institute of Automation, Chinese Academy of Sciences \\
$^2$University of Chinese Academy of Sciences, Beijing, China \\
$^3$Department of Electrical and Computer Engineering, National University of Singapore \\
\{jiangyan.yi, ruibo.fu, jhtao, shuai.nie, haoxin.ma\}@nlpr.ia.ac.cn, haizhou.li@u.nus.edu}

\begin{document}
%
\maketitle
\begin{abstract}
Audio deepfake detection is an emerging topic, which was included in the ASVspoof 2021. However, the recent shared tasks have not covered  many real-life and challenging scenarios.
The first Audio Deep synthesis Detection challenge (ADD) was motivated to fill in the gap.
The ADD 2022 includes three tracks: low-quality fake audio detection (LF), partially fake audio detection (PF) and
audio fake game (FG). The LF track focuses on dealing with bona fide and fully fake utterances with various real-world
noises etc. The PF track aims to distinguish the partially fake audio from the real. The FG track is a rivalry game, which includes
two tasks: an audio generation task and an audio fake detection task. In this paper, we describe the datasets, evaluation
metrics, and protocols. We also report major findings that reflect the recent advances in audio deepfake detection tasks.
The ADD 2022 dataset is publicly available, see Train\&Dev\footnote{https://zenodo.org/records/12188127}, Adaption\footnote{https://zenodo.org/records/12188083}, Track1 eval\footnote{https://zenodo.org/records/10843991}, Track2 eval\footnote{https://zenodo.org/records/12187997}, Track3.2 R1 eval\footnote{https://zenodo.org/records/12188035}, Track3.2 R2 eval\footnote{https://zenodo.org/records/12188055}.

\end{abstract}

\section{Introduction}
\label{sec:intro}

Over the last few years, the technology of speech synthesis and voice conversion \cite{Wang2017Tacotron,Shen2018Wavenet, Wang2018Style, Wang2021Prosody} has made significant improvement with the development of deep learning. The models can generate realistic and human-like speech. It is difficult for most people to distinguish the generated audio from the real. However, the technology also poses a great threat to the society if some attackers misuse it. Therefore, a lot of efforts have been made for audio deepfake detection task recently \cite{Wang2012Eval, Chen2020Gen, Wang2020Deep,Yi2021Half,Ma2021Continual}.

The ASVspoof challenges have been organized to detect spoofed audio for automatic speaker verification systems. The ASVspoof 2015 \cite{Wu2015ASVspoof} involves logical access (LA) task detecting synthetic and converted speech. The ASVspoof 2017 \cite{Kinnunen2017ASVspoof} only includes replay attacks named physics access (PA) task. The ASVspoof 2019 \cite{Todisco2019ASVspoof} consists of two tasks: LA and PA. There are three tasks in the ASVspoof 2021 \cite{2021ASVspoof}: LA, PA and speech deepfake (DF). The ASVspoof challenges have played a key role in fostering spoofed speech detection research, which mainly aim to protect automatic speaker verification systems from manipulation. Although the audio deepfake detection task is included in the ASVspoof 2021 \cite{2021ASVspoof}, it only involves compressed audio similar to the LA task. However, it ignores many challenging attacking situations in realistic scenarios. (1) Diverse background noises and disturbances are contained in the fake audios. (2) Several small fake clips are hidden in a real speech audio. (3) New algorithms of speech synthesis and voice conversion are proposed rapidly. These pose a serious threat since that it is difficult to deal with the above-mentioned attacking situations.

Therefore, we launched the first Audio Deep synthesis Detection challenge (ADD 2022) to fill in the gap. It includes three tracks, which consider some challenging fake situations in real life. We hope that the ADD 2022~\footnote{http://addchallenge.cn} can spur researchers around the world to build innovative new technologies that can further accelerate and foster research on detecting deepfake and manipulated audios.

The rest of this paper is organized as follows. Section 2 describes tracks of the ADD 2022. Datasets and evaluation metrics are introduced in Section 3 and 4. Section 5 presents the detection baseline models and challenge results. This paper is concluded in Section 6.
\vspace{-10pt}
\section{Tracks}
\label{sec:format}

The ADD 2022 challenge includes three tracks: low-quality fake audio detection (LF), partially fake audio detection (PF) and audio fake game (FG).

\textbf{Track 1. LF}: It focuses on dealing with bona fide and fully fake utterances with various real-world noises and background music effects etc. The fake audios are generated using a large variety of text-to-speech and voice conversion algorithms.

\textbf{Track 2. PF}: It aims to distinguish the partially fake audio from the real. The partially fake utterances generated by manipulated the original bona fide utterances with real or synthesized audio \cite{Yi2021Half}.

\textbf{Track 3. FG}: It includes two tasks: an audio generation task and an audio fake detection task.

\textbf{Track 3.1 generation task (FG-G)}: It aims to generate fake audios that can fool the fake detection model in track 3.2. The participants are encouraged to generate attack samples according to the given text and speaker identities, and the generated attack samples should reach certain intelligibility and similarity.

\textbf{Track 3.2 detection task (FG-D)}: It tries to detect all the fake audios, especially the attack samples generated from track 3.1. There are two rounds of evaluations in track 3.2. The first round evaluation data contains a set of unseen genuine and deepfake audios. The second round evaluation data contains some generated speech utterances submitted by track 3.1.

The goal of track 1 and 2 is to develop a method or an algorithm to distinguish the generated audio from the bona fide. Track 3 is a rivalry game for participants to generate adversarial samples and improve the anti-attack ability of the detection model from two sides \cite{Peng2021DFGC}. Participants in this track can choose to either create adversarial samples to attack the detection model as much as possible, or improve the anti-attack ability of the detection model.
\vspace{-10pt}
\section{Datasets}
\label{sec:pagestyle}

The datasets for the challenge consist of training, dev, adaptation and test sets. All the tracks use the same training and dev sets. Different adaptation and test sets are provided for each track. There is no speaker overlap among training, dev, adaptation and test sets. The training, dev and adaptation sets are provided with both input and ground truth. The test sets are provided without ground truth. Some utterances are selected from Mandarin publicly available corpus AISHELL-1 \cite{aishell_2017}, AISHELL-3 \cite{Shi2020aishell3} and AISHELL-4 \cite{2021AISHELL4} to build the datasets. The statistics of datasets provided by the ADD 2022 challenge are reported in Table \ref{tab:train} and \ref{tab:test}.

\vspace{-10pt}
\subsection{Training and dev sets}
The training and dev sets include genuine and fake utterances. The datasets are based upon a large-scale and high-fidelity multi-speaker Mandarin speech corpus called AISHELL-3.
40 male speakers and 40 female speakers are selected from AISHELL-3 corpus to build the training and dev sets. The set of speakers is partitioned into two speaker-disjoint sets for training and dev. The genuine utterances of the training and dev sets are selected from the AISHELL-3. The mainstream speech synthesis and voice conversion systems are used to generate the fake audios. For track 3.1, participants are recommended to build a multi-speaker speech synthesis or voice conversion based on the AISHELL-3.

\vspace{-10pt}
\subsection{Adaptation sets}
The adaptation set of each detection task is provided for the participants. There are three adaptation sets.

\textit{Track 1}: It is composed of genuine and fully fake utterances contained various noises.

\textit{Track 2}: It consists of partially fake utterances generated by manipulated the original genuine utterances with real or synthesized audios.

\textit{Track 3.2}: It includes various fake audios generated by the organizers. The given speaker identity and content are synthesized by the speech synthesis systems provided by organizers.

\vspace{-10pt}
\subsection{Test sets}
The test sets include unseen genuine and fake utterances. Three test sets are provided for the participants.

\textit{Track 1}: It is composed of unseen genuine and fully fake utterances with various noises.

\textit{Track 2}: It consists of unseen genuine and partially fake utterances.

\textit{Track 3.1}: 10 speakers ID from AIShell-3 dataset are listed as the evaluation speaker ID.

\textit{Track 3.2}: The first round (R1) test set is similar to track 1. The second round (R2) test set includes the R1 test set and some of generated speech audios submitted by track 3.1.

\begin{table}[t]
	\caption{The utterances of training, dev. and adaptation sets.}
	\label{tab:train}
	\centering
	\begin{tabular}{cccccc}
		\toprule
		\multirow{2}{*}{ } & \multirow{2}{*}{Training} & \multirow{2}{*}{Dev.} & \multicolumn{3}{c}{ Adaptation} \\ \cmidrule(l){4-6}
		&               &        & Track 1  &  Track 2   & Track 3.2     \\ \midrule
		Genuine &   3012     & 2307         &  300     & 0     & 0    \\
		Fake &   24072   & 26017            &   700   & 1052      & 893      \\ \bottomrule
	\end{tabular}
\end{table}

\begin{table}[t]
	\caption{The statistics of test sets for detection tasks.}
	\label{tab:test}
	\centering
	\begin{tabular}{ccccc}
		\toprule
		\multirow{2}{*}{Test} & \multirow{2}{*}{Track 1} & \multirow{2}{*}{Track 2} & \multicolumn{2}{c}{ Track 3.2} \\ \cmidrule(l){4-5}
		&            &         & R1 &  R2 \\ \midrule
		\#Utterances & 109199 & 100625 & 112861 & 126861   \\ \bottomrule
	\end{tabular}
\end{table}

\vspace{-10pt}
\section{Evaluation metrics}
\label{sec:typestyle}
The goal of track 1, 2 and 3.2 is to develop a method or an algorithm to distinguish the generated audio from the real. So equal error rate (EER) \cite{Wu2015ASVspoof} is used as the evaluation metric for these tracks. The generation task in track 3.1 aims to generate fake audios that can fool the fake detection model in track 3.2. Therefore, the evaluation metric of track 3.1 is the deception success rate (DSR).

\vspace{-10pt}
\subsection{Equal error rate (EER)}
Previously, EER is used by Wu et al. in the ASVspoof challenge \cite{Wu2015ASVspoof}.
The metric for ADD 2022 is the 'threshold-free' EER, defined as follows. Let $P_{fa}(\theta)$ and $P_{miss}(\theta)$ denote the false alarm and miss rates at threshold $\theta$.

\vspace{-10pt}
\begin{align}\label{eer}
P_{fa}(\theta)&=\frac{\# \{{\textit{fake trials with score $\textgreater$ $\theta$}} \}}{\# \{{\textit{total fake trials}} \}} \\
P_{miss}(\theta)&=\frac{\# \{{\textit{genuine trials with score $\textless$ $\theta$}} \}}{\# \{{\textit{total genuine trials}} \}}
\end{align}
\vspace{-10pt}

So $P_{fa}(\theta)$ and  $P_{miss}(\theta)$ are, respectively, monotonically decreasing and increasing functions of $\theta$.  The EER corresponds to the threshold $\theta_{EER}$ at which the two detection error rates are equal, i.e. $EER=P_{fa}(\theta_{EER}) =P_{miss}(\theta_{EER})$.

There are two rounds of evaluations in track 3.2. Each round evaluation have each own ranking in terms of EER. The final ranking is in terms of the weighted EER (WEER), which is defined as follow.
\vspace{-5pt}
\begin{eqnarray}\label{weer}
WEER= \alpha * EER\_{R1} + \beta * EER\_{R2}
\end{eqnarray}
where $\alpha = 0.4$ and $\beta = 0.6$, $EER\_{R1}$ and $EER\_{R2}$ are the EER of R1 and R2 evaluation in track 3.2, respectively.

\begin{table}
\caption{Description of detection baseline systems}
\label{tab:baselines}
\centering
\begin{tabular}{cccc}
\toprule
 ID & Model & Features & Training data \\\midrule
S01 & GMM &LFCC & Training set \\
S02 & GMM &LFCC & Training and adaptation sets \\
S03 & LCNN &LFCC & Training set \\
S04 & LCNN &LFCC & Training and adaptation sets \\
S05 & RawNet2 &Raw & Training set \\
S06 & RawNet2 &Raw & Training and adaptation sets \\
\bottomrule
\end{tabular}
\end{table}

\vspace{-10pt}
\subsection{Deception success rate (DSR)}
Track 3 is a rivalry game for participants to generate adversarial samples and improve the anti-attack ability of the detection model from two sides. Therefore, Track 3.1 and 3.2 are evaluated separately. The deception success rate (DSR) and ERR are chosen as the metric for track 3.1 and 3.2, respectively.
DSR reflects the degree of fooling the audio deepfake detection model by the generated utterances, which is defined as followed:
\vspace{-8pt}
\begin{eqnarray}\label{dsr}
DSR=\frac{W}{A*N}
\end{eqnarray}
where $W$ is the count of wrong detection samples by all the detection models on the condition of reaching each own EER performance, $A$ is the count of all the evaluation samples, and $N$ is the number of detection models.

To avoid cheating by submitting interference samples, the intelligibility and similarity are also evaluated by multiple methods. Each submitted sample should meet the text and speaker information requirements of the competition.

\begin{table}
\caption{Results are in terms of EER (\%) for Track 1.}
\label{tab:track1}
\centering
\begin{tabular}{ccc|ccc|ccc}
\toprule
\# & ID & EER & \# & ID & EER & \# & ID & EER \\\midrule
1 & A01 & 21.7  & 17 & A15 & 28.0  & 33 & \textbf{S03} & 32.3 \\
2 & A02 & 23.0  & 18 & A16 & 28.2  & 34 & A30 & 32.8 \\
3 & A03 & 23.8  & 19 & A17 & 28.4  & 35 & A31 & 32.8 \\
4 & \textbf{S02} & 24.1  & 20 & A18 & 29.2  & 36 & A32 & 33.0 \\
5 & \textbf{S01} & 25.2  & 21 & \textbf{S04} & 29.9  & 37 & A33 & 33.8 \\
6 & A04 & 25.9  & 22 & A19 & 29.9  & 38 & \textbf{S06} & 33.9 \\
7 & A05 & 26.1  & 23 & A20 & 30.0  & 39 & A34 & 34.0 \\
8 & A06 & 26.3  & 24 & A21 & 30.2  & 40 & \textbf{S05} & 35.2 \\
9 & A07 & 26.6  & 25 & A22 & 30.6  & 41 & A35 & 35.9 \\
10 & A08 & 26.8  & 26 & A23 & 30.6  & 42 & A36 & 37.7 \\
11 & A09 & 26.8  & 27 & A24 & 31.0  & 43 & A37 & 41.2 \\
12 & A10 & 27.1  & 28 & A25 & 31.7  & 44 & A38 & 41.2 \\
13 & A11 & 27.3  & 29 & A26 & 32.0  & 45 & A39 & 42.9 \\
14 & A12 & 27.3  & 30 & A27 & 32.0  & 46 & A40 & 43.6 \\
15 & A13 & 27.4  & 31 & A28 & 32.1  & 47 & A41 & 46.2 \\
16 & A14 & 27.9  & 32 & A29 & 32.2  & 48 & A42 & 67.1 \\
  &   &   &   &   &   &   & Avg. & 31.7 \\
\bottomrule
\end{tabular}
\end{table}

\section{Challenge results}
\label{sec:majhead}
Participants could submit detection scores and receive results by CodaLab website. The datasets were requested by more than 120 teams from 15 countries for all tracks.

\begin{table}
\caption{Results are in terms of EER (\%) for Track 2.}
\label{tab:track2}
\centering
\begin{tabular}{ccc|ccc|ccc}
\toprule
\# & ID & EER & \# & ID & EER & \# & ID & EER \\\midrule
1 & B01 & 4.8 & 12 & B12 & 36.3 & 23 & B22 & 46.3 \\
2 & B02 & 7.9 & 13 & B13 & 38.6 & 24 & \textbf{S02} & 47.5 \\
3 & B03 & 9.4 & 14 & B14 & 38.6 & 25 & \textbf{S03} & 47.8 \\
4 & B04 & 16.6 & 15 & B15 & 39.4 & 26 & \textbf{S04} & 48.1 \\
5 & B05 & 20.6 & 16 & B16 & 40.5 & 27 & B23 & 50.0 \\
6 & B06 & 25.6 & 17 & B17 & 40.5 & 28 & \textbf{S05} & 50.1 \\
7 & B07 & 26.0 & 18 & B18 & 40.8 & 29 & \textbf{S06} & 50.2 \\
8 & B08 & 30.6 & 19 & B19 & 40.9 & 30 & B24 & 50.6 \\
9 & B09 & 34.6 & 20 & B20 & 42.5 & 31 & B25 & 54.0 \\
10 & B10 & 34.7 & 21 & B21 & 42.9 & 32 & B26 & 55.8 \\
11 & B11 & 35.4 & 22 & \textbf{S01} & 45.8 & 33 & B27 & 57.0 \\
  &   &   &   &   &   &   & Avg. & 37.9 \\
\bottomrule
\end{tabular}
\end{table}

\begin{table}
\caption{Results are in terms of DSR (\%) for Track 3.1.}
\label{tab:track3.1}
\centering
\begin{tabular}{ccc|ccc|ccc}
\toprule
\# & ID & DSR & \# & ID & DSR & \# & ID & DSR \\\midrule
1 & C10 & 93.8  & 6 & C15 & 54.6  & 11 & C08 & 37.8 \\
2 & C05 & 91.6  & 7 & C01 & 52.7  & 12 & C09 & 36.6 \\
3 & C14 & 89.5  & 8 & C13 & 49.0  & 13 & C03 & 29.1 \\
4 & C02 & 72.4  & 9 & C07 & 41.0  & 14 & C11 & 25.6 \\
5 & C04 & 72.4  & 10 & C12 & 39.6  &   & Avg. & 56.1 \\
\bottomrule
\end{tabular}
\end{table}
\begin{table*}[ht]
\caption{Results are in terms of EER (\%) for R1 and R1 evaluation, and WEER(\%) for final evaluation in Track 3.2.}
\label{tab:track3.2}
\centering
\begin{tabular}{ccccc|ccccc|ccccc}
\toprule
\# & ID & EER\_R1 & EER\_R2 & WEER & \# & ID & EER\_R1 & EER\_R2 & WEER & \# & ID & EER\_R1 & EER\_R2 & WEER \\\midrule
1 & D01 & 8.6  & 11.1  & 10.1  & 14 & \textbf{S02} & 12.2  & 19.4  & 16.5  & 27 & D21 & 10.6  & 100.0  & 64.2 \\
2 & D02 & 9.4  & 11.0  & 10.4  & 15 & D13 & 20.1  & 14.2  & 16.6  & 28 & D22 & 15.7  & 100.0  & 66.3 \\
3 & D03 & 8.3  & 12.1  & 10.6  & 16 & D14 & 13.9  & 18.4  & 16.6  & 29 & D23 & 100.0  & 44.3  & 66.6 \\
4 & D04 & 9.6  & 12.0  & 11.0  & 17 & D15 & 16.2  & 17.1  & 16.7  & 30 & D24 & 19.0  & 100.0  & 67.6 \\
5 & D05 & 8.6  & 12.8  & 11.1  & 18 & \textbf{S01} & 14.1  & 19.3  & 17.2  & 31 & D25 & 19.7  & 100.0  & 67.9 \\
6 & D06 & 8.5  & 13.4  & 11.4  & 19 & D16 & 15.1  & 19.9  & 18.0  & 32 & D26 & 20.3  & 100.0  & 68.1 \\
7 & D07 & 8.8  & 13.4  & 11.6  & 20 & \textbf{S03} & 18.6  & 17.6  & 18.0  & 33 & D27 & 21.3  & 100.0  & 68.5 \\
8 & D08 & 8.8  & 13.7  & 11.7  & 21 & \textbf{S06} & 16.7  & 22.6  & 20.2  & 34 & D28 & 21.6  & 100.0  & 68.6 \\
9 & D09 & 9.6  & 14.3  & 12.4  & 22 & \textbf{S05} & 20.6  & 22.1  & 21.5  & 35 & D29 & 23.7  & 100.0  & 69.5 \\
10 & D10 & 12.0  & 15.0  & 13.8  & 23 & D17 & 20.7  & 27.8  & 25.0  & 36 & D30 & 24.2  & 100.0  & 69.7 \\
11 & D11 & 11.9  & 15.2  & 13.9  & 24 & D18 & 24.5  & 26.0  & 25.4  & 37 & D31 & 26.5  & 100.0  & 70.6 \\
12 & D12 & 14.4  & 13.7  & 14.0  & 25 & D19 & 27.9  & 24.3  & 25.7  & 38 & D32 & 27.6  & 100.0  & 71.0 \\
13 & \textbf{S04} & 11.5  & 17.3  & 15.0  & 26 & D20 & 100.0  & 13.4  & 48.0  & 39 & D33 & 28.2  & 100.0  & 71.3 \\
 &  &  &   &   &   &  &   &  &  &   & Avg. & 20.7 & 43.1 & 34.2 \\
\bottomrule
\end{tabular}
\vspace{-8pt}
\end{table*}
\subsection{Detection baselines}

ADD 2022 adopted six detection baseline systems.
Motivated by the ASVspoof challenge \cite{2021ASVspoof}, we use Gaussian mixture model (GMM), light convolutional neural network (LCNN) \cite{Wu2020LCNN} and RawNet2 \cite{Jung2020Rawnet2} to train baseline models.
We modified the officially released source code~\footnote{http://github.com/asvspoof-challenge/2021} to build GMM, LCNN and RawNet2 classifiers. The input features of GMM and LCNN models are linear frequency cepstral coefficients (LFCCs) \cite{Sahi2015A}. Raw audio waveforms are used as the input of RawNet2 models.

All baseline models were trained using only the respective ADD 2022 training data or adaptation data. They were optimised using only the respective development (Dev.) data. None used any kind of data augmentation. The description of the six baselines are listed in Table \ref{tab:baselines}.

\vspace{-10pt}
\subsection{Results and analysis}

Table \ref{tab:track1}, \ref{tab:track2} and \ref{tab:track3.2} shows results in terms of EER for track 1, 2 and 3.2.  The results in terms of DSR for track 3.1 are reported in Table \ref{tab:track3.1}.

For LF task, the average EER of all submissions is 31.7\% and the best result shows an detection EER of 21.7\%. Only 3 of the 42 participating teams produced systems that outperformed the best baseline S02. The GMM baseline model achieved the lowest EER compared with LCNN and RawNet2 baseline models. All the baselines obtained performance gains, when the model trained with training and adaptation sets directly.

For PF task, the average EER of all submissions is 37.9\% and the best result was 4.8\% in term of EER. The performance of the best baseline S01 was bettered by 21 of the 27 participating teams. The GMM baseline model also achieved the best result. However, all the baselines obtained worse performance, when the model trained with training and adaptation sets directly.

For FG-D task, the final average WEER of all submissions is 34.2\% and the final lowest WEER of 10.1\%. The average EER of all submissions is 20.7\% and the lowest EER of 8.3\% in the R1 evaluation. The average EER of all submissions is 43.1\% and the lowest EER of 11.0\% in the R2 evaluation. The EER of 100.0\% denotes that the result was not submitted by the participant.
For FG-G task, there are 14 teams submitted the generated audios. The best DSR was achieved by 93.8 \% , and the average DSR was 56.1 \%.

It is still challenging for all tracks, especially for LF track. Although the best result achieves by 4.8\% in terms of EER, the average EER is still high for PF track. When adding some generated fake samples from FG-G task into the evaluation dateset,the performance of FG-D task degrades obviously.

\vspace{-10pt}
\section{Conclusions}
\label{sec:page}
This paper summarises the challenge task, datasets, preliminary evaluation results and analysis.
ADD 2022 addressed three different challenging fake scenarios, namely LF, PF and FG, involving four tasks. The results show that it is difficult to use the same model to deal with all fake scenarios. The result also show that detection generalisation remains an open problem. The detection model will be fooled easily with low quality and unseen generated fake utterances. Whether the evaluation metrics is reasonable or not is needed to discuss further. So generalisation and evaluation metrics will remain a focus for future evaluations.

\vspace{-10pt}
\section{ACKNOWLEDGMENTS}
\label{sec:illust}

This work is supported by the National Key Research and Development  Plan of China (No.2020AAA0140003), the National Natural Science Foundation of China (NSFC) (No.61901473, No.62101553, No.61831022).

\bibliographystyle{IEEEbib}
\bibliography{strings,refs}

\end{document}